\documentclass[aps,prd,superscriptsize,preprint,longbibliography,nofootinbib]{revtex4-1}
\usepackage[utf8]{inputenc} 

\def\letter{0}
\def\letter{1}
\pdfoutput=1 

\usepackage{epsfig}
\usepackage{epstopdf}
\usepackage{textcomp}
\usepackage{graphicx}
\usepackage{ifthen}
\usepackage{ulem}

\usepackage{amsmath}
\usepackage[psamsfonts]{amssymb}
\usepackage{euscript}

\usepackage{latexsym}
\usepackage[arrow,matrix,curve]{xy}

\usepackage[hypertexnames=false]{hyperref}
\hypersetup{
    colorlinks=true,
    linkcolor=blue,
    filecolor=magenta,   
    citecolor=black,   
    urlcolor=cyan,
    }

\newskip\humongous \humongous=0pt plus 1000pt minus 1000pt

\newif\ifdtup

\def\,{\hspace{-.1cm}}

\def\fc#1#2 {\frac{n}{q}#1\frac{n}{q}#2}

\renewcommand{\cos}{\textrm{cos}}
\renewcommand{\sin}{\textrm{sin}}

\renewcommand{\tanh}{\textrm{tanh}}
\newcommand{\sech}{\textrm{sech}}

\renewcommand{\theequation}{\arabic{section}.\arabic{equation}}
\renewcommand{\(}{\begin{equation}}
\renewcommand{\)}{end{equation} \vspace{-.05in}\linebreak}

\newcounter{saveeqn}
\newcounter{savealpheqn}

\newcommand{\alpheqn}{\setcounter{saveeqn}{\value{equation}}%
  \stepcounter{saveeqn}\setcounter{equation}{0}%
  \renewcommand{\theequation}{\mbox{\arabic{section}.\arabic{saveeqn}
\alph{equation}}}
  \renewcommand{\)}{\end{equation}}}
\def\part#1{\frac{\partial}{\partial{#1}}}%
\def\group#1{\refstepcounter{equation}\setcounter{saveeqn}
 {\value{equation}}%
  \label{#1}\setcounter{equation}{0}%
\renewcommand{\theequation}{\mbox{\arabic{section}.\arabic{saveeqn}
\alph{equation}}}
  \renewcommand{\)}{\end{equation}}}
\newcommand{\reseteqn}{\setcounter{equation}{\value{saveeqn}}%
  \renewcommand{\theequation}{\arabic{section}.\arabic{equation}}%
  \renewcommand{\)}{\end{equation}}}

\newcommand{\aalpheqn}{\setcounter{saveeqn}{\value{equation}}%
  \stepcounter{saveeqn}\setcounter{equation}{0}%
  \renewcommand{\theequation}{\mbox{
        \Alph{subsection}.\arabic{saveeqn}\alph{equation}}}
   \renewcommand{\)}{\end{equation}}}
\newcommand{\areseteqn}{\setcounter{equation}{\value{saveeqn}}%
  \renewcommand{\theequation}{\Alph{subsection}.\arabic{equation}}%
  \renewcommand{\)}{\end{equation}}}

\renewcommand{\(}{\begin{equation}}
\renewcommand{\)}{\end{equation}}
\newcommand{\ba}{\begin{eqnarray}}
\newcommand{\ea}{\end{eqnarray}}
\newcommand{\cbp}{\mathop{\vtop{\ialign{##\crcr
   $\hfil\displaystyle{}\hfil$\crcr\noalign{\kern-13pt\nointerlineskip}
   \BIG{)}\hskip0pt\crcr\noalign{\kern3pt}}}}}
\newcommand{\pa}{\mathop{\vtop{\ialign{##\crcr

$\hfil\displaystyle{\oplus}\hfil$\crcr\noalign{\kern+1pt\nointerlineskip
}
   \hspace{.08in}$^{\alpha=0}$\hskip6pt\crcr\noalign{\kern3pt}}}}}

\newcommand{\pp}{^{\prime\prime}}

\catcode`\@=11
\def\vereq#1#2{\lower3pt\vbox{\baselineskip1.5pt \lineskip1.5pt
\ialign{$\m@th#1\hfill##\hfil$\crcr#2\crcr\sim\crcr}}}
\catcode`\@=12

\renewcommand{\(}{\begin{equation}}
\renewcommand{\)}{\end{equation}}

\usepackage[dvipsnames]{xcolor}

\newcommand{\beas}{\begin{eqnarray*}}
\newcommand{\eeas}{\end{eqnarray*}}

\newcommand{\bquo}{\begin{quote}}
\newcommand{\enqu}{\end{quote}}

\def\lim#1{\stackrel{\rm{lim}}{{}_{#1}}}

    \newcommand{\g}{\mathfrak g}

\def\XXint#1#2#3{{\setbox0=\hbox{$#1{#2#3}{\int}$}
     \vcenter{\hbox{$#2#3$}}\kern-.5\wd0}}

\newcommand{\beq}{\begin{equation}}
\newcommand{\eeq}{\end{equation}}
\newcommand{\bea}{\begin{eqnarray}}
\newcommand{\eea}{\end{eqnarray}}

\newskip\humongous \humongous=0pt plus 1000pt minus 1000pt

\newif\ifdtup

\catcode`\@=11

\ifthenelse{\equal{\letter}{0}}{ 

\@addtoreset{equation}{section}
\def\theequation{\arabic{section}.\arabic{equation}}

\def\@normalsize{\@setsize\normalsize{15pt}\xiipt\@xiipt
\abovedisplayskip 14pt plus3pt minus3pt%
\belowdisplayskip \abovedisplayskip
\abovedisplayshortskip \z@ plus3pt%
\belowdisplayshortskip 7pt plus3.5pt minus0pt}

\def\small{\@setsize\small{13.6pt}\xipt\@xipt
\abovedisplayskip 13pt plus3pt minus3pt%
\belowdisplayskip \abovedisplayskip
\abovedisplayshortskip \z@ plus3pt%
\belowdisplayshortskip 7pt plus3.5pt minus0pt
\def\@listi{\parsep 4.5pt plus 2pt minus 1pt
      \itemsep \parsep
      \topsep 9pt plus 3pt minus 3pt}}

\relax

\def\section{\@startsection{section}{1}{\z@}{3.5ex plus 1ex minus  .2ex}{2.3ex plus .2ex}{\large\bf}}

\def\thesection{\arabic{section}}
\def\thesubsection{\arabic{section}.\arabic{subsection}}

\def\appendix{\setcounter{section}{0}
 \def\thesection{Appendix \Alph{section}}
 \def\thesubsection{\Alph{section}.\arabic{subsection}}
 \def\theequation{\Alph{section}.\arabic{equation}}}
\renewcommand{\theequation}{\arabic{section}.\arabic{equation}}

}{
\renewcommand{\theequation}{\arabic{equation}}

} 

\begin{document}

\title[
]{The Universal Floquet Modes of (Quasi)-Breathers and Oscillons}

\author{Jarah Evslin}
\email{jarah@impcas.ac.cn}

 \affiliation{Institute of Modern Physics, NanChangLu 509, Lanzhou 730000, China}
\affiliation{University of the Chinese Academy of Sciences, YuQuanLu 19A, Beijing 100049, China}

\author{Tomasz Roma\'nczukiewicz}
\email{tomasz.romanczukiewicz@uj.edu.pl}

\affiliation{ Institute of Theoretical Physics,  Jagiellonian University, Lojasiewicza 11, 30-348 Krak\'{o}w, Poland}

\author{Katarzyna Slawi\'nska}
\email{katarzyna.slawinska@uj.edu.pl}

\affiliation{ Institute of Theoretical Physics,  Jagiellonian University, Lojasiewicza 11, 30-348 Krak\'{o}w, Poland}

\author{Andrzej Wereszczynski }
\email{andrzej.wereszczynski@uj.edu.pl}

\affiliation{
 Institute of Theoretical Physics,  Jagiellonian University, Lojasiewicza 11, 30-348 Krak\'{o}w, Poland
}

\affiliation{International Institute for Sustainability with Knotted Chiral Meta Matter (WPI-SKCM2), Hiroshima University, Higashi-Hiroshima, Hiroshima 739-8526, Japan}

\begin{abstract}
\noindent
Just as linearized perturbations of time-independent configurations can be decomposed into normal modes, those of periodic systems can be decomposed into Floquet modes, which each evolve by a fixed phase over one period.  We show that in the case of a (1+1)-dimensional relativistic field theory with a single scalar of mass $m$, all breathers, quasi-breathers and oscillons of length $1/\epsilon$ have identical nonrelativistic Floquet modes at leading order in an $\epsilon/m$ expansion.  More precisely, these Floquet modes depend only on $\epsilon$ and $m$, and are independent of the potential of the theory.  In particular, there is a continuum of Floquet modes corresponding to each real momentum plus four discrete modes corresponding to space translations, time translations, boosts and amplitude changes.  There are no discrete shape modes.  We provide simple, explicit formulas for these universal leading-order, nonrelativistic Floquet modes.  
\end{abstract}

\maketitle

\section{Introduction}

Field theories describing a single  mass $m$ scalar field $\phi$ subjected to a potential $V(\phi)$ often enjoy breather, quasi-breather or oscillon solutions \cite{Bogolyubsky:1976nx, Gleiser:1993pt, Copeland:1995fq} in which the field, in a region of size $1/\epsilon$, oscillates about some minimum of the potential. Not surprisingly, the properties of the oscillons depend on the details of the model, see e.g. \cite{Amin:2011hj,Salmi:2012ta,Fodor:2006zs, Olle:2019kbo, Olle:2020qqy, vanDissel:2023zva, Blaschke:2024uec}. However, despite this large variety, if $\epsilon\ll m$ then at leading order in $\epsilon/m$ the shapes of these solutions are universal \cite{Fodor:2008es}, \footnote{Whether such a solution is a breather or not depends on the choice of potential. If it is not, whether it is a quasi-breather or an oscillon depends on the choice of boundary conditions. However, the difference in boundary conditions vanishes to all orders in $\epsilon$ and so does not affect our leading order in $\epsilon$ results. We remark that this universality does not concern all known oscillons. Especially, there are oscillons in theories without the mass threshold, $m=0$ \cite{Dorey:2023sjh, Blaschke:2024dlt, vanDissel:2025xqn}.}.  The amplitude of the oscillation is proportional to $\epsilon$ with a constant of proportionality $\lambda_F$ that depends on the potential.

What are the linearized perturbations about such a solution?  One might expect that they will depend on $m$, $\epsilon$, the amplitude of the oscillation and the details of the potential.  Below we will show that in 1+1 dimensions, in the case of normal modes with wave numbers well below $m$, the dependence on the amplitude and on the details of the potential cancel one another at leading order in the dimensionless $\epsilon/m$, so that these nonrelativistic linearized perturbations depend only on the two dimensionful quantities $m$ and $\epsilon$.  

We will use this observation as follows.  The exact perturbations of the Sine-Gordon breather are in principle known, as a result of the integrability of the Sine-Gordon model.  We will extract the linear order of the nonrelativistic oscillations.  These are the Floquet modes of the Sine-Gordon breather.  However, as the Floquet modes are universal, these will be the nonrelativistic Floquet modes of all 1+1 dimensional (quasi)-breathers and oscillons.  Indeed, we will note that these are solutions of the coupled sets of ordinary differential equations derived for such Floquet modes in Ref.~\cite{Evslin:2024sup}.  The relativistic Floquet modes, on the other hand, have already been found analytically in Ref.~\cite{Evslin:2024sup}.  They are not universal but depend on a single parameter.

\section{Small oscillons and their Floquet modes}
We will consider a 1+1 dimensional classical field theory with a scalar field $\phi(x)$ and its conjugate momentum $\pi(x)$ subjected to a Hamiltonian
\beq
H=\int dx \left[\frac{\pi(x)^2+\partial_x\phi(x)\partial_x\phi(x)}{2}+\frac{V(g\phi(x))}{g^2}\right]
\eeq
where $g$ is a coupling constant.  We will demand that $\phi=0$ be a local minimum of $V$. This includes the Sine-Gordon model and also many popular models of oscillons such as the $\phi^3$ model and the $\phi^4$ double-well.  

We will define the mass $m$ to be the square root of the second derivative of $V$ evaluated at $\phi=0$ and more generally we will let $V^{(n)}$
be its $n$th derivative at $\phi=0$.  Define the effective coupling $\lambda_F$ by 
\beq
\lambda_F=\frac{5V^{(3)\ 2}}{6m^2}-\frac{V^{(4)}}{2}.
\eeq
Then it is well-known \cite{Fodor:2008es} that, if $\lambda_F>0$, then for every {\it small} nonzero $\epsilon$ with dimension of mass, there is a breather, quasi-breather or oscillon solution of the classical equations of motion $\phi(x,t)=f(x,t)$ where $f(x,t)$ is given by \footnote{One should be aware that {\it this is not} the necessary condition for the existence of oscillons and oscillons which violate this condition exist \cite{Blaschke:2024dlt}.}
\begin{equation}
f(x,t)=\frac{\epsilon}{g\sqrt{2\lambda_F}}\sech(\epsilon x)\cos(\Omega t)
+\epsilon^2\frac{2V^{(3)}}{3g^2\lambda_F m^2}\sech^2(\epsilon x)\left(\cos(2\Omega t)-3\right)
+O(\epsilon^3/m^3). \label{osc}
\end{equation}
The fundamental frequency of the oscillon is assumed to be close to the mass threshold
\begin{equation}
\Omega=\sqrt{m^2-\epsilon^2}+O(\epsilon^4/m^3).
\end{equation}
This solution depends on the inverse length $\epsilon$, the mass $m$ and also on the third and fourth derivatives of the potential.
Here we have expanded in powers of the dimensionless combination $\epsilon/m$.  

Now let us consider a perturbation $\g(x,t)$ so that
\beq
\phi(x,t)=f(x,t)+\g(x,t).
\eeq
Then, up to linear order in $\g(x,t)$, $\phi(x,t)$ is a solution to the classical equations of motion if
\bea
(\partial_t^2-\partial^2_x+m^2)\g(x,t)&=&-\left[ 
\frac{\epsilon V^{(3)}}{\sqrt{2\lambda_F}}\sech(\epsilon x)\cos(\Omega t)
\right.\label{pad}\\
&&\left. \hspace{-2cm}
+
\frac{2\epsilon^2}{\lambda_F}\sech^2(\epsilon x)\left[
\left(\frac{V^{(3)\ 2}}{3m^2}+V^{(4)}\right)\cos(2\Omega t)+V^{(4)}-\frac{V^{(3)\ 2}}{m^2}
\right]
+O(\epsilon^3/m)
\right]\g(x,t).\nonumber
\eea
The present letter concerns the solutions of this equation.  Note that the equation itself does not appear at all universal, with explicit dependence of $V^{(3)},\ V^{(4)}$ and the combination $\lambda_F$. 

Let us consider a Floquet mode
\beq
\g(x,t+2\pi/\Omega)=e^{-i 2\pi \omega/\Omega}\g(x,t). \label{flo}
\eeq
Now restrict attention to the nonrelativistic modes by letting
\beq
\omega=\omega_2\epsilon^2
\eeq
where $\omega_2$ is of order $O(1/m)$.  These modes are nonrelativistic \footnote{ This condition is actually more stringent than we need, as, the results below will follow so long as $\omega/m\ll 1$.} because $\omega$ is of order $O(\epsilon^2/m)$ and we consider $\epsilon\ll m$.


We will decompose the normal modes in powers of $\epsilon$
\beq
\g(x,t)=\sum_{j=1}^\infty \epsilon ^j \g_j(x,t)
\eeq
and further decompose $\g_1(x,t)$ as
\beq
\g_1(x,t)=G(\epsilon x)e^{-i(\Omega+\omega)t}+H(\epsilon x)e^{i(\Omega-\omega)t}.
\eeq
This automatically satisfies Eq.~(\ref{flo}).  One might add multiples of $\Omega$ to the exponent, but then it would not satisfy Eq.~(\ref{pad}) at order $O(\epsilon)$.

At order $O(\epsilon^2)$, Eq.~(\ref{pad}) can always be satisfied by appropriately choosing $\g_2(x,t)$.  However, at order $O(\epsilon^3)$, the coefficients of $e^{-i(\pm\Omega+\omega)t}$ in $\g_3$ are annihilated by the left hand side, and so must also vanish on the right hand side.  Physically, this condition imposes that the modes are not resonant.  The conditions that these two coefficients vanish are respectively~\cite{Evslin:2024sup}
\bea
-(1+2m\omega_{2})H_k(\epsilon x)+ H\pp_k(\epsilon x)+2\sech^2 (\epsilon(x-x_0))(G_k(\epsilon x)+2H_k(\epsilon x))&=&0\label{esp}\\
(-1+2m\omega_{2})G_k(\epsilon x)+ G\pp_k(\epsilon x)+2\sech^2 (\epsilon(x-x_0))(2G_k(\epsilon x)+H_k(\epsilon x))&=&0.\nonumber
\eea
Surprisingly, these equations are independent of the potential. As a consequence, the Floquet modes themselves do not depend on the particularities of the model. 

As the solutions are independent of the potential, we are free to choose any potential we wish. We have therefore chosen the case of the Sine-Gordon breather.  In this case, integrability allows the perturbations to be calculated exactly.  These perturbations were calculated in Appendix C of Ref.~\cite{Dashen:1975hd} which stated that they were obtained using the Backlund transformation of Ref.~\cite{Hirota}.  We have obtained \footnote{The form in Eqs. (C8) and (C9) of Ref.~\cite{Dashen:1975hd} contains several typos.  Also, it uses complex values of several parameters that Ref.~\cite{Hirota} states should be real, and an analytic continuation that has some ambiguities due to branch cuts.  We corrected the typos before deriving the Floquet modes.} the linearized normal modes by linearizing these results, leading to the universal solutions
\bea
G_k(\epsilon x)&=&\left({\sech^2(\epsilon x)+2m\omega_{k,2}-2}{}\right)\ \cos\left(\sqrt{2m\omega_{k,2}-1}\epsilon x\right)\nonumber\\
&&-{2\sqrt{2m \omega_{k,2}-1}\ \tanh(\epsilon x)\sin\left(\sqrt{2m\omega_{k,2}-1}\epsilon x\right)}{}\nonumber\\
H_k(\epsilon x)&=&{\sech^2(\epsilon x)}{}\ \cos\left(\sqrt{2m\omega_{k,2}-1}\epsilon x\right)
\eea
for the even modes and by
\bea
G_k(\epsilon x)&=&\left({\sech^2(\epsilon x)+2m\omega_{k,2}-2}{}\right)\sin\left(\sqrt{2m\omega_{k,2}-1}\epsilon x\right)\nonumber\\
&&+{2\sqrt{2m\omega_{k,2}-1}\ \tanh(\epsilon x)}{}\cos\left(\sqrt{2m\omega_{k,2}-1}\epsilon x\right)
\nonumber\\
H_k(\epsilon x)&=&{\sech^2(\epsilon x)}{}\ \sin\left(\sqrt{2m\omega_{k,2}-1}\epsilon x\right)
\eea
for the odd modes.  Here we have added the index $k$ which we will use to label the solutions and their Floquet coefficients.  Here $k$ is a continuous parameter and so these are the continuum modes.  Thus, while these indeed solve (\ref{esp}), they do not exhaust the solutions.

Note that $H$ has support inside the oscillon, while $G$ oscillates asymptotically with a wavenumber of $\sqrt{2m\omega_{k,2}-1}\epsilon$. Thus, one can formally treat these modes as half-bound modes or Feschbach resonances \cite{Feshbach}. Interestingly, such modes play a significant role in dynamics of solitons, see e.g., their participation in interaction of monopoles \cite{Forgacs:2003yh}, vortices \cite{Bachmaier:2025igf}, kinks \cite{GarciaMartin-Caro:2025zkc} and $Q$-balls \cite{Ciurla:2024ksm}. 

We see that $\omega_{k,2}\geq 1/(2m)$ and that the wavenumber is $\pm\epsilon\sqrt{2m\omega_{k,2}-1}$. If we identify $k$ with this wavenumber, then the Floquet coefficient is
\beq
\omega_k=\epsilon^2\omega_{k,2}=\frac{\epsilon^2+k^2}{2m}
\eeq
and $\Omega+\omega_k$ is the usual frequency $\sqrt{m^2+k^2}$.  In other words, with $\omega_{k,2}$ of order $O(1/m)$, the wavelength of the perturbation is of order the size of the oscillon itself.  As $m\omega_{k,2}$ grows to be much larger than unity, $H/G$ is inversely proportional to $m\omega_{k,2}$, and so for the high energy modes, $H$ can be ignored.  In the case of relativistic modes, for which $m\omega_{k,2}$ becomes of order $O(m^2/\epsilon^2)$ or equivalently $\omega_k\sim m$, the time derivative of (\ref{pad}) appears already at the leading order in our $\epsilon$ expansion and so the modes above no longer solve (\ref{pad}) at order $O(\epsilon)$.  In fact, in this case the Floquet modes were already found explicitly in Ref.~\cite{Evslin:2024sup} and they are not universal. 

The universal solutions above have two very nice properties.  If $(G_{k_1},H_{k_1})$ and $(G_{k_2},H_{k_2})$ are two such solutions, then they are orthogonal in the sense
\beq
\int dx (G_{k_1}(\epsilon x) G_{k_2}(\epsilon x)- H_{k_1}(\epsilon x) H_{k_2}(\epsilon x))=C_{k_1}2\pi\delta(k_1-k_2) \label{ort}
\eeq
with a normalization constant $C_{k}=2m^2\omega_{k,2}^2/(\epsilon\sqrt{2m\omega_{k,2}-1})$.  Second, 
\beq
\int dx \left( G_{k_1}(\epsilon x)H_{k_2}(\epsilon x)-G_{k_2}(\epsilon x)H_{k_1}(\epsilon x)\right)=0.
\eeq
In the quantum theory, these two relations will be used to show that the annihilation operators for various normal modes commute, and so all of the Floquet modes can be simultaneously placed in their ground states.

In addition to the continuum modes, there are also four discrete Floquet modes, corresponding to $\omega=0$
\bea
\g_B(\epsilon x,t)&=& {\rm{tanh}}\left(
\epsilon x
\right) {\rm{sech}}\left( 
\epsilon x
\right)\cos\left( \Omega t\right)\\
\g_T(\epsilon x,t)&=&{\rm{sech}}\left( 
\epsilon x
\right)\sin\left( \Omega t\right)\nonumber\\
\g_M(\epsilon x,t)&=&t\g_B(\epsilon x,t)+x\g_T(\epsilon x,t)\nonumber\\
\g_{\epsilon}(\epsilon x,t)&=&-x{\rm{sech}}\left(
\epsilon x
\right){\rm{tanh}}\left( 
\epsilon x
\right)\cos\left( \Omega t\right).\nonumber
\eea
These four perturbations correspond to infinitesimal translations along the four dimensional moduli space of the classical solutions (\ref{osc}).  In particular, the first corresponds to a spatial translation, the second to a time translation, the third to a boost, and the last to a change in the amplitude or equivalently the thickness $\epsilon$. These four Floquet modes are also universal, as they depend only on the solution inverse size $\epsilon$ and also on the mass $m$ which determines the frequency $\Omega$.

\section{Conclusions}

In the present paper, we showed that in the long wavelength limit the Floquet modes of the small (quasi)-breather or oscillons possess universal exact form. These are the zero modes and continuum modes. 

Note that there are no discrete nonzero-frequency bound modes in this regime. However, perturbed oscillons often reveal a non-trivial structure of isolated well-defined peaks in the power spectrum, some of them inside the gap, indicating that there are bound-like excitations \cite{Blaschke:2025anm, Alonso-Izquierdo:2025iet}. In fact, the linearization of the oscillon leads to an infinite ladder of components with frequencies $\Omega+n\rho$, where $n \in \mathbb{Z}$. Some of them can be located below the mass threshold, but the rest are propagating in the continuum. Thus, a mode should be viewed as a sort of Feshbach resonance, i.e., a partially-bound mode with some components bounded to the soliton (frequencies below the mass threshold) and with some components propagating in the continuum. 

This does not contradict our findings. The observed universality applies when the wavelength is longer than $1/m$ which, in terms of frequency, is equivalent to when the difference between the oscillon frequency and the mode frequency is small. This difference itself it seems also has a gap $\sim \epsilon^2/2m$ \cite{Evslin:2024sup}, and so the claim is that there are no shape modes in that gap. This agrees with the results presented here and in \cite{Evslin:2024sup}.

\section*{Acknowledgement}

\noindent
This work was supported by the Higher Education and Science Committee of the Republic of Armenia (Research Project No. 24RL-1C047). K. S. acknowledges
financial support from the Polish National Science
Centre (Grant No. NCN 2021/43/D/ST2/01122).

\bibliography{Bibliography.bib}

\end{document}